\documentstyle[12pt]{article}
\begin{document}
\begin{center}
{\Large
{\bf Off-Shell Rho-Omega Mixing Through Quark Loops With Non-Perturbative
 Meson Vertex And Quark Mass Functions}}\\
\vspace{1.cm}
{A. N. Mitra$^{a,b}$ and K.-C. Yang$^a$}\\
\vspace{0.5cm}
{$^a$Dept. of Physics, National Taiwan University,
Taipei,Taiwan-10764,R.O.C}\\
\vspace{1.cm}

{$^b$Dept. of Physics, University of Delhi, Delhi 110007, India}\\
\end{center}
\vspace{1cm}
%
%\date{\today}

\begin{abstract}
The momemtum dependence of the off-shell $\rho$-$\omega$ mixing amplitude
is calculated through a two-quark loop diagram, using non-perturbative
meson-quark vertex functions for the $\rho$ and $\omega$ mesons, as well
as non-perturbative quark propagators. Both these quantities are generated
self-consistently through an interlinked BSE-cum-SDE approach with a 3D
support for the BSE kernel with two basic constants which are pre-
checked against a wide cross section of both meson and baryon spectra
within a common structural framework for their respective 3D BSE's. With
this pre-calibration, the on-shell strength works out at
-2.434$\delta(m_q^2)$	in units of the change in
"constituent mass squared", which is consistent with the $e^+e^-$ to
$\pi^+\pi^-$ data for a u-d mass difference of ~4 MeV ,while the relative
off-shell strength (0.99 $\pm$ 0.01) lies midway between
quark-loop and QCD-SR results.
We also calculate the photon-mediated $\rho$-$\omega$ propagator whose
off-shell structure has an additional pole at $q^2$=0. The implications
of these results vis-a-vis related investigations are discussed.
\end{abstract}

%\narrowtext

During the last couple of years, the problem of charge symmetry violation
(CSV) in nuclear physics \cite{1}, especially a possible resolution of the
so-called Nolen-Schiffer anomaly (NSA) \cite{2} by the ${\rho_0-\omega}$
mixing contribution to the N-N force \cite{3},has come in for a good deal
of attention \cite{4}-\cite{8}. The renewed interest was triggered by the
possibility of an off-shell momentum dependence \cite{4} in the ${\rho-\omega}$
mixing amplitude which would be expected to arise through any reasonable form
of dynamics
generated by an intermediate fermion-antifermion loop,wheher at the quark
level (u-d mass difference) \cite{4} or at the nucleon level (n-p mass
difference) \cite{7}. Since the original suggestion \cite{4}, several more
quark level calculations \cite{5,6,8} have appeared for a quantitative
handle on the issue of momentum dependence of the ${\rho-\omega}$ \cite{6,8}
and the analogous ${\pi-\eta}$ \cite{5} amplitudes, using respectively the
results of the effective chiral Lagrangians \cite{9,10}, the field-
theoretic language of Schwinger-Dyson Equation for the quark propagator
\cite{11}, and the methodology \cite{12,13} of QCD sum rules \cite{13}.
All these calculations reveal in varying degrees the effect of off-shell
momentum dependence of the ${\rho-\omega}$ amplitude on the physics of the
N-N potential in its sensitive region. The estimates  however
seem to vary over a wide range, from relatively gentle variations via the
quark loop \cite{4,6} or nucleon loop \cite{7} mechanisms, to rather large
effects \cite{8} through the QCD sum rule method, though none of these
methods seem to be free from uncertainties of input parametrizations. The
results are expressed by a certain output parameter ${\lambda}$ which
is roughly
a measure of the degree of off-shellness of the ${\rho-\omega}$ mixing
amplitude ${\theta (q^2)}$ defined on  a linear scale through its
dependence on the variable ${q^2}$ \cite{5,8}
(but adapted to the euclidean metric) as follows :
\begin{equation}
\theta ( q^2)=\theta (M^2)[1 - \lambda(1 +( q^2/ M^2)]
\label{1}
\end{equation}
where ${ M}$ is the mean ${\rho-\omega}$ mass and the magnitude of
${\lambda}$ is the quantity of main dynamical interest \cite{4,5,6,7,8}.

\section{Quark-Loop Method in 4D vs. 3D BSE}
In the QCD-SR method, the inputs are ${< q \bar q>}$, ${\delta<q \bar q>}$
and the current ${ u, d}$ masses which are more or less under control.
However the uncertainties arise from the 'matching' of the two sides of
the 'duality' equation by the 'standard' methods \cite{12,13} to obtain
the necessary stability in the sum-rule structure, which is the key to
the determination of the hadronic parameter ${\lambda}$ \cite{8}.

In the quark-loop method \cite{4,6}, on the other hand, the dynamics resides
in the hadron-quark vertex function and constituent quark mass \cite{4},
or better, the dynamical mass function \cite{6}, and these two quantities
in turn are interlinked through the Schwinger-Dyson and Bethe-Salpeter
Equations, a formalism which, though well known since the seventies
\cite{11},has been greatly revived in the nineties \cite{14,15}. These
approaches also require in practice a generous degree of parametrization
\cite{14,15} of the basic entities, since any exact solution of these
coupled equations is still a distant dream. Therefore any attempt to
adapt this methodology to the present problem \cite{6} must also share
the corresponding parametric uncertainties \cite{15}, without prior checks
from other sectors of hadron physics, notably spectroscopy as well as
some sensitive coupling constants and form factors. A broad perspective on
these approaches \cite{14,15} has been discussed more fully elsewhere
\cite{16},in the context of an alternative program with a similar philosophy
\cite{17}-\cite{20} whose continued emphasis on the spectroscopy sector
stems from its sensitivity to the 'gluon-like propagator' in the infrared
region' (a paraphrase for the '${ q-\bar q}$ potential' in the more
ordinary language).

We shall not go into the relative merits of the QCD-SR \cite{12} and
the BSE-cum-SDE methods, which have been discussed elsewhere \cite{16}.
However in the context of the latter it will be useful to make a
distinction between two broad types, the
'spectroscopy-oriented' type \cite{16} $\to$ \cite{20} which depends
on a basically 3D approach, and the more "orthodox" 4D
type \cite{14,15}. For purposes of this paper we shall term these two
BSE-cum-SDE methods as "3D" and "4D" BSE's respectively for short. Both
make use of the two-quark loop integration to calculate the rho-omega
amplitude but  this difference in terminology emphasises the  difference
in the parametrizations of the
infrared region of the gluonic propagator  which are 3D and 4D respectively.
While the 4D form is prima facie more natural, the theoretical reasons for
the 3D form are no less persuasive and the interested reader will find the
necessary details on its theoretical motivations \cite{22} from various
angles in \cite{16,20}. Here we shall merely cite the chief expreimental
reason, viz., the ${ O(3)}$-like spectra in the PDG-tables \cite{23}
continually for four decades which provides the bed-rock of foundation for
any theoretical effort at a microscopic description of quark structure, and
our 3D BS-programme \cite{16}-\cite{21} has been specifically designed to
meet this requirement.On the other hand, a literal consequence of the 4D
form of
parametrization of the infrared part (confining) of the gluon propagator
\cite{15} would be to predict ${ O(4)}$-like spectra which contradicts
experiment \cite{16}. The reason why such ${O(4)}$-like results are not
entirely visible in some of these spectral predictions \cite{15} is merely
because of their consideration of mainly the ground state masses ${L=0}$,
since the fuller implications of the 4D forms would not start showing up
until the predictions include the ${L}$-excited states \cite{16}.

Since some quark loop results of the 4D-BSE \cite{15} are already available
\cite{6}, along with those of the QCD-SR analysis \cite{8}, it should be
of considerable interest, even without prejudice to the question
of ${O(3)}$ versus ${O(4)}$-like spectra (important as it may be
in its own right), to record for comparison the corresponding results
of the 3D-BSE \cite{16}, in view of their parameter-free
nature. This may be useful in the context of the current controversy
\cite{4,6,7,8} on the
off-shell strength of the ${\rho-\omega}$ mixing amplitude which is an
important parameter for charge symmetry breaking in the $(N-N)$ force.
We recall in this
connection that the 3D-BSE formalism is specifically calibrated to both
${q-\bar{q}}$ \cite{18} and ${qqq}$ \cite{19} spectra, both in excellent
accord with data \cite{23}, as well as to a representative list of hadron
couplings \cite{24,16,17}. All this has been obtained with just two basic
constants ${C_0,\omega_0}$ ${common}$ to both types, since a third
input, the quark mass
(constituent), gets dynamically generated thruogh the (chiral-symmetry
breaking) solution of the SDE \cite{11,16}, so that in this
(spectroscopy-oriented) BSE-cum-SDE approach there is practically no  scope
for any free parametrization beyond the ones noted above, a condition
which is probably important for the determination of the rather sensitive
parameters ${\theta(M^2)}$ and ${\lambda}$ under study.

\section{3D BSE-cum-SDE Formalism}
The purpose of this paper is to present the results of this calculation
in the most economical fashion, omitting all but the essential details.
To that end we shall use \cite{8} for the definitions and notations of
the crucial parameters involved, and calibrate our language to that of
\cite{4,6} as far as possible for the definition of the loop integrals,
except for the implicit understanding of an euclidean metric notation
underlying our formulation, and the use of ${P}$ for their ${q}$ notation
since the latter has in all our formulations \cite{16,17,20} stood for the
internal 4-momentum of the quarks within the hadron, while ${P}$ is the
4-momentum of the (composite) hadron. As to the 3D-BSE formalism itself we
shall make free use of \cite{16,17}, but recapitulate some essential
results so as to keep the paper within	reasonably self-contained limits.

The quantities we shall explicitly calculate in our formalism are :
(i) the function ${\theta (P^2)}$ with an explicit proportionality to
${\delta(m_q^2)}$ where ${m_q^2}$ is the constituent (dynamical) quark
mass 'squared', which is obtained directly from an analytical formula
(given below) for ${\Pi(P^2)}$ \cite{8}, by  a simple process of
differentiation w.r.t. ${m_q^2}$ ;
(ii) the parameter ${\lambda}$ \cite{8} which can be explicitly identified
from the linear dependence of this quantity on the inverse meson propagator
${(P^2+M^2)}$;
(iii) the analogous ${\rho-\omega}$ potential mediated by an intermediate
photon, so that its full off-shell structure is a chain of two linear
off-shell quantities ${g_{\rho-\gamma}}$ and ${g_{\omega-\gamma}}$ to be
compared with the ${\theta}$-function ,eq,(1), for ${\rho-\omega}$
mixing due to the "strong" effect of ${u,d}$ difference which involves
this linear factor in the off-shell quantity ${(P^2+M^2)}$ only once.
(Despite the comparative weakness of the e.m. effect, its off-shell scenario
is, on this account, somewhat different from that of the (strong) ${u-d}$
effect; this point is discussed further at the end).

As to the actual numerical values,  the only unknown quantity
in our formalism is the ${u-d}$ mass difference  which we shall keep
as a free multiplicative factor for $\theta(M^2)$, to facilitate
discussion at the end on this point. The main results are:
\begin{equation}
\theta(M^2) = - 1289(MeV) \cdot \delta(m_q) ;  \lambda = 0.99 \pm 0.01
\label{2}
\end{equation}
where for ${\delta(m_q)}$ it is usual \cite{6} to take ${m_d - m_u}$.

Our formalism is based on the Covariant Instaneity Ansatz (CIA) \cite{17}
which gives the Bethe-Salpeter Kernel ${K(q,q')}$ for a quark-antiquark
system a 3D support expressed through its dependence on the component of
${q_\mu}$ transverse to ${P_\mu}$, for which we use a hat notation,i.e.,
\begin{equation}
\hat q_\mu = q_\mu - q\cdot P P_\mu / P^2
\label{3}
\end{equation}
so that ${K(q,q')}$ = ${K(\hat q, \hat q')}$ for a 3D support \cite{17}.
As a result of this ansatz, there is an exact interconnection between
the 3D and 4D forms of the BSE \cite{17} and the hadron-quark vertex
function ${\Gamma_H(q,P)}$ becomes a function  ${\Gamma_H(\hat q)}$
of a single argument ${\hat q_\mu}$. It is usually convenient to take
out the Dirac matrix from this structure, viz.,${\gamma_5}$ for a pion,
${i \gamma \cdot \rho}$ for a rho-meson, etc.; the multiplying scalar
factor ${\Gamma_h(\hat q)}$ which carries the dynamical information
has the following universal structure \cite{17} :
\begin{equation}
\Gamma_h(\hat q) = N_H\times D(\hat q)\otimes \phi(\hat q)/(2\pi)^{5/2}
\label{4}
\end{equation}
Here $D(\hat q)$ is a 3D denominator function, and $\phi(\hat q)$
the corresponding wave function which together satisfy a Lorentz-covariant
Schroedinger-like equation of the form $D\phi = \int K\phi$,
representing the 3D reduction of the 4D BSE as a result of the above
ansatz. The quantity $N_H$ is the standard 4D BS normalizer which goes with
the vertex function (4). The $D$-function for equal mass kinematics has
the simple form
\begin{equation}
D(\hat q) = 4\omega (\omega^2 - M^2/4); \omega^2 = m_q^2 + \hat{q}^2 ,
\label{5}
\end{equation}
while the $\phi$-function is model dependent. In particular, a gaussian form
\begin{equation}
\phi(\hat q) = exp(-\hat{q}^2/(2\beta^2))
\label{6}
\end{equation}
emerges (for harmonic confinement) as a solution of the 3D BSE, with
$\beta^2$ obtained $analytically$ from the input structure of the BS-kernel
\cite{16,18} and checked against Spectroscopy \cite{23}. Its value for the
$\rho-\omega$ case is $0.0692$ \cite{18}. In eq.(5),
$M$ is the hadron mass and $m_q$ the constituent (dynamical) quark mass.
Its momentum dependence was obtained in \cite{16} by simply relating the
quark mass function to the pion-vertex function which must reduce to each
other in the chiral limit of vanishing pion mass (${M_\pi = 0}$),
by virtue of the Ward-Takahashi identity for the axial-vector vertex
function \cite{11}. Therefore by specializing eqs.(4-6) to the pion case
in the limit ${M_\pi = 0}$, one immediately obtains the formula \cite{16} :
\begin{equation}
m(\hat{p})= m_q^{-2}(m_q^2+\hat{p}^2 )^{3/2}\cdot exp(-\hat{p}^2/2\beta_\pi^2)
\label{7}
\end{equation}
where the quantity $\beta_\pi^2$ (=0.031 for the pion case \cite{18}) is
still governed by the same BS-dynamics \cite{16,18}, but now (because of
the goldstone nature of the pion in the chiral limit) the normalization
has had to be fixed anew by identifying the "constituent" mass $m_q$ with
this function at its zero-momentum limit (${m(0)=m_q}$). In terms of
$m(\hat p)$ the non-perturbative quark propagator $S_F(p)$ is now given by
\begin{equation}
(S_F(p))^{-1} = i ( m(\hat p) + i \gamma\cdot p )
\label{8}
\end{equation}
where the Landau gauge is understood ($A(p^2)=1$,\cite{14,16}) and
$m(\hat p)$ is given by (7). This non-perturbative mass function was employed
in \cite{16} for evaluating the quark condensate $<q\bar q>$ as an
explicit quadrature :
\begin{eqnarray}
< q\bar q > = \frac{6}{\pi^2} \int d^3\hat p \frac{m(\hat p)}{\sqrt
{\hat p^2 + m(\hat p)^2}}
\nonumber
\end{eqnarray}
giving a value in the QCD-SR range \cite{12}; the meaning of
$\hat p_\mu$ was also clarified therein.

An important property of the structure (5) for the quantity $D(\hat q)$
in the hadron-quark vertex function is that it prevents,
through a general cancellation mechanism \cite{17,20}, the occurrence of
overlapping pole effects due to integration over the time-like component
of the loop-momentum in any  quark-loop integral, and thus automatically
pre-empts the possibility of any  "free" propagation of quarks that might
otherwise occur. It thus may be regarded as a simple 3D alternative to the
construction of quark propagators as entire functions through  more
elaborate models \cite{15,6}, but
with the added benefit of a parameterfree description (c.f.,\cite{6}).
This structure will play a key role in simplifying the loop-integral
for the meson self energy operator from its 4D scalar form, eq.(10), to
the 3D form, eq.(12), as given below.

\section{The $\rho \to \omega$ Amplitude}
After collecting these essential ingredients of the "3D" BSE formalism,
we now turn to the central quantity, viz.,the two-loop contribution to
the meson self-energy operator $\Pi_{\mu\nu}(P^2)$ \cite{4,6,8} which is
expressible as
\begin{equation}
\rho_\mu \Pi_{\mu\nu}(P^2) \omega_\nu = i (2\pi)^4\times
\int d^4q {\Gamma_h(\hat q)}^2 Tr[i\gamma\cdot\rho S_F(q+P/2) i\gamma
\cdot\omega S_F(-q+P/2)]
\label{9}
\end{equation}
where $\Gamma_h(\hat q)$ is the scalar part of the vertex function defined
by eq.(4), and the $\rho$,$\omega$ symbols on both sides of (9) stand for
their respective polarization vectors. At this stage the scalar vertex
function is common to $\rho$ and $\omega$, since their mass difference due to
the $u-d$ effect will be automatically taken care of via standard
differentiation w.r.t. $m_q^2$, c.f.\cite{4};(see below). Simplifying the
trace in eq.(9) and checking on current conservation (which is routinely
satisfied) we can write ${\Pi_{\mu\nu}(P^2)}$ as ${(\delta_{\mu\nu}-P_\mu
P_\nu\cdot P^{-2}) \Pi(P^2)}$ where, following any one of \cite{17,20,25},
\begin{equation}
\Pi(P^2) =  2i(2\pi)^4 N_V^2 \int d^4q [D^2(\hat q) \phi^2(\hat q)\cdot
(\Delta_1+\Delta_2-P^2-4\hat q^2/3)/(\Delta_1 \Delta_2)]
\label{10}
\end{equation}
where \cite{17,20,25}
\begin{equation}
\Delta_{1,2} = m_q^2 + (P/2 \pm q)^2 ; (P^2=-M^2)
\label{11}
\end{equation}
The integration over the longitudinal (time-like) component of $q_\mu$,
viz., $M d\sigma$, ($\sigma$ equals $q \cdot P/P^2$), is carried out again
as in \cite{17,20,25} wherein the structure (5) of the $D$ function
ensures an
exact cancellation of the effects of overlapping singularities arising from
the $\sigma$-pole residues. The resultant 3D integration over $d^3\hat q$
is expressible as
\begin{equation}
\Pi(P^2) = - 2 N_V^2 \int d^3\hat q \phi^2(\hat q)[D^2(\hat q)/\omega
- D(\hat q)\cdot (P^2 + 4\hat q^2/3)].
\label{12}
\end{equation}

Eq.(12) brings out explicitly, without further ado, the linear
structure of the mass operator in the off-shell variable $P^2$. The BS
normalizer $N_V^2$ in eq.(12) is itself an integral of the same kind
as $\Pi(P^2)$, and is formally defined for any V-meson through the
equation \cite{24,17}:
\begin{equation}
2i P\mu N_V^{-2} = (2\pi)^4 \int d^4q{\Gamma_h(\hat q)}^2 Tr[i\gamma\cdot V
S_F(q + P/2)i\gamma_\mu S_F(q + P/2)i\gamma\cdot V S_F(-q + P/2)]
\label{13}
\end{equation}
whose integral over the time-like component of $q\mu$ can be carried out
exactly as above to give a formula analogous to eq.(12) :
\begin{equation}
N_V^{-2} = 2\int d^3\hat{q}\phi^2(\hat{q})4\omega\times
[\omega^2-\hat q^2/3] =  0.0502 (GeV)^{-6}
\label{14}
\end{equation}

The quantities $\omega$ and $D$ in both equations (12,13) are
defined as in eq.(5) which in turn carries the explicit information
on the $m_q^2$ dependence of both these quantities. This fact
facilitates a simple
differentiation w.r.t. $m_q^2$ under the integral signs in eqs.(12,13)
in order to evaluate $\delta[\Pi(P^2)]$ which precisely represents,
with no further normalization, the desired quantity $\theta(P^2)$
defined in eq.(1), while the values of its two crucial parameters as
predicted by this model are already listed in eq.(2). In obtaining
the latter we have used the equality of $\delta(m_q^2)$ with
2$m_q\cdot \delta(m_q)$, and  employed the 'spectroscopic' value
265 Mev for $m_q$, the constituent mass \cite{18,19}. For the evaluation
of the integrals (12,13) we have not explicitly considered the momentum
variation (7) of the dynamical mass, but left it at its 'constituent'
value $m_q$  corresponding to zero momentum. This has been done mainly
for simplicity and transparency in carrying out the differentiation
process. Though not strictly  valid. the scope of error on
this account is likely to be small for two reasons : (i) the main burden
of momentum variations in the two integrals (12,13) is carried by the
meson-quark vertex function whose effect has been fully incorporated via
eqs.(4-6); (ii) the mass function, eq(7), maintains a sort of plateau
(250-300 Mev) in the region of integration which provides the
bulk contributions to the  integrals. Our estimate of error, based on
some trial runs with the momentum-dependent mass function, is about 10$\%$.
On the other hand the explicit analytic structures in $m_q^2$ of the
integrals (12,13) greatly minimize the possibility of further numerical
errors that would be inherent in the differentiation process in the
absence of a $(non-perturbative)$ analytical form which is usually more
difficult to ensure than, e.g.,a point vertex structure \cite{4}, without
additional parametric assumptions on the way, e.g., \cite{6}.

Before comparing our results with others, we wish to record for
completeness the predictions of this model on the photon-mediated chain
of ${\rho-\gamma-\omega}$ mixing amplitude which we denote by
$\theta_\gamma(P^2)$ in the same relative normalization as eq.(1).
Here we need no longer distinguish between $m_u$ and $m_d$ and take a
simple proportionality of the ${\rho-\gamma}$ and ${\omega-\gamma}$
amplitudes to a common dynamical quantity $g_V(P^2)$ defined by
\begin{equation}
g_V(P^2) V_\mu = -i\int d^4q {\Gamma_h(\hat q)}\times
Tr[i\gamma\cdot V S_F(q+P/2) i\gamma_\mu S_F(-q+P/2)]/\sqrt 2,
\label{15}
\end{equation}
the multiplicity factors being $e$ and $e$/3 respectively, and $V_\mu$
standing collectively for $\rho$ or $\omega$.
The other symbols are as defined in eq.(9) and earlier. The evaluation of
$g_V(P^2)$ is on lines similar to eq.(9),but actually simpler, and leads
to the explicit formula
\begin{equation}
g_V(P^2) = 4\sqrt(3/2) \beta^3 N_V [2m_q^2+4\beta^2-(P^2+M^2)/2]
\label{16}
\end{equation}
Writing it in a forn analogous to eq.(1), we have
\begin{equation}
g_V(P^2) = f_V(M^2)[1 - \mu (1 + P^2/M^2)] ;
\label{17}
\end{equation}
where the on-shell value $f_V(M^2)$ and the off-shell coefficient $\mu$ are
\begin{equation}
g_V(M^2) = 0.1608(Gev)^2 ; \mu = M^2/[4(m_q^2 + 2\beta^2)] = 0.7197.
\label{18}
\end{equation}
The final result for the complete photon-mediated $\rho$-$\omega$
amplitude is
\begin{equation}
\theta_\gamma(P^2) =  \frac{e^2}{3} g_V(P^2)\frac{1}{-P^2} g_V(P^2),
\label{19}
\end{equation}
where we have explicitly shown the photon propagator in the middle, to
bring out the 'extended' nature of the off-shell extrapolation
due to the photon-mediated mixing compared to that due to the ${u-d}$
effect, despite the smallness of (19) compared to (1). Unlike
(1), there is no uncertainty in (19) within this model, though the
on-shell value $(P^2=-M^2)$ is a bit too high (see discussion below).
\begin{equation}
\theta_\gamma(M^2)= +1316(Mev)^2
\label{20}
\end{equation}
The off-shell effect, on the other hand, is best expressed through the
corresponding $N$-$N$ potentials \cite{8} which are given by
\begin{equation}
V(\rho-\omega) = - [\theta(M^2)/2M][1-(2\lambda/Mr)]exp(-Mr);
\label{21}
\end{equation}
\begin{equation}
V(\rho-\gamma-\omega) =[\theta_\gamma(M^2)^2/M^4]\cdot
 [(i-\mu)^2/r + [(2\mu-1)/r]exp(-Mr) - (M/2)exp(-Mr)]
\label{22}
\end{equation}
respectively, where a common factor $g_{\rho N}$ $g_{\omega N}$ \cite{8}
has been suppressed from the last two equations. Eq.(21) has no
counterpart of the ${1/r}$ term in (22).

\section{Results and Discussion}
To put the results of this investigation in perspective with those in
\cite{4}$\to$\cite{8} we should first note that in this spectroscopy-rooted
approach there is little scope for any significant variation of the
input parameters ($\omega_0$, $C_0$, and $m_q$) whose respective values
(158 Mev, 0.27, and 265 Mev)
can be traced all the way back to the BS-Kernel itself \cite{18},
without effecting a simultaneous change in the (already good) fits \cite{18}
to the observed meson spectra \cite{23}, and in the more recent (equally
good) fits \cite{19} to the baryon spectra \cite{23} with these very
parameters. It is with this constraint that the numbers obtained above
may be viewed vis-a-vis those in \cite{4}$\to$\cite{8}, especially in
respect of the off-shell parameter $\lambda$, eq(1), which can be compared
with almost all of them. However, the on-shell value $\theta(M^2)$ ,eq(2),
is quite specific in this model, and could at at best be compared with
the predictions of, say, chiral Lagrangian models \cite{9,10}, except
that the available prediction \cite{5} refers to $\pi^0$-$\eta$ mixing
and cannot be used for a direct comparison.

The only uncertainty in our on-shell value, eq.(2), arises from a
corresponding uncertainty in the value of $\delta(m_q)$ for which a natural
substitute, $a la$ Politzer \cite{26}, would be $(m_d-m_u)$. The latter
quantity has been discussed in great detail in \cite{8} to which we
refer the interested reader, but for a definitive estimate it should be
reasonable to take a value, say, 4 Mev \cite{6} which is well within the
limits of the \cite{8} analysis. With this value we get $\theta(M^2)$
equal to (-5156)$MeV^2$ which should be compared with the value
(-4520 $\pm$ 600) obtained from $e^+e^-$ $\to$ $\pi^+\pi^-$ data \cite{27},
after taking account of the ${\rho-\gamma-\omega}$ chain which gives a
smaller contribution of opposite sign, viz., +1316, eq.(20). Its
inclusion  gives the net value -3840 which is still within the
experimental range \cite{27}, (taking account of the uncertainties of
the u-d mass difference).

The somewhat larger value of $\theta_\gamma(M^2)$ compared to the
"VMD" value \cite{28} 610 $MeV^2$ quoted in \cite{8} may in turn be related
to the quantity $g_V(M^2)$,eq.(18), which gives 0.1608 $GeV^2$. This number,
when divided by $M_\rho$ = 0.775 GeV, precisely translates,in the QCD-SR
\cite{29} notation, to the result $f_\rho$ = 215 MeV, to be compared to the
quoted value of 200 MeV \cite{29} needed for agreement with the
$\rho$$\to$$e^+e^-$ width. This is the extent of our overestimate of
$\theta_\gamma(M^2)$ compared to the VMD-value \cite{28,8}, but nevertheless
tolerable enough to warrant a discussion (below) of the off-shell aspects
of $\rho$-$\gamma$-$\omega$ mixing along with that of the main (u-d) term.

The off-shell prediction is dominated by the parameter $\lambda$, eq.(1)
at the value (0.99), eq.(2), and its photonic counterpart $\mu$ defined
in eq.(18) at the value (0.720). Our value
of $\lambda$ is rather below the QCD-SR range (1.43-1.85) \cite{8},
implying a "softer" off-shell effect in this quark-loop
model than the "harder" effect in the QCD-SR approach, as already noted
in \cite{8} for QCD-SR versus quark-loop methods: A smaller value of
$\lambda$ would tend to postpone the onset of
attenuation of the ${\rho-\omega}$ mixing potential due to the off-shell
effects, to somewhat shorter distances, as measured by
the "critical distance" \cite{8}  $r_c$ = 2$\lambda_M$, which is also
seen from eq.(21). In a similar way, the off-shell effect of the
photon-mediated ${\rho-\omega}$ mixing, as measured by the parameter
$\mu$ = 0.720,eq.(18), produces the potential,eq.(22), but its $(1/r)$-term
has no counterpart in eq.(21). Taking note of the $opposite$ signs of the
two effects, the following scenario emerges: The  two short range terms
of (21) get duly reduced
by the two corresponding terms of (22) by about 20-25 $\%$.
 However, the long range
$(1/r)$-term of (22), which has no counterpart in (21), reinforces the
$exp(-Mr)$ term of the latter, again by about 20-25$\%$ near the critical
distance \cite{6,8}, but continues with increasing strength down to
shorter distances and therefore further postpones the attenuation by
another (small) amount. For brevity we omit further discussion \cite{8}.

Finally we wish to comment on the magnitude of our $\lambda$-value,
0.99 $\pm$ 0.01
vis-a vis other determinations \cite{4,5,6,7,8}.

We have checked on the
possible variation in this quantity due to the (neglected) effect of the
momentum dependence of the dynamical mass eq.(7), and found the effect
to be $\leq$ 10$\%$. There is little scope
for further variation in this otherwise 'rigid' description, unless a
totally different set of input parameters ( $C_0$, $\omega_0$,$m_q$ )
from the ones \cite{18,19} employed here, produces an equally good fit
\cite{18,19} to the observed spectra \cite{23}, which is rather unlikely.
Nevertheless this value seems to lie about  midway between
other quark-loop calculations \cite{4,5,6} and	QCD-SR results \cite{8},
though somewhat nearer to the former than to the latter; rather surprisingly,
it is quite close to the nucleon-loop value \cite{7} of about unity
\cite{8}.
It is also in fair agreement with the corresponding results \cite{5} for
$\pi^0$-$\eta$ mixing obtained from  chiral Lagrangian models \cite{9,10},
though a similar result for the analogous case of
$\rho$-
$\omega$ mixing by the same method \cite{9,10} is not yet available. Of course
a ${non-linear}$
dependence of $\theta(P^2)$ on $P^2$, such as attempted in \cite{6},
may well change this (linear) scenario, but this requires more effort.

To summarise, we have outlined an explicit calculation of the $\rho$-
$\omega$ mixing amplitude, both on- and off- shell, in the form expressed
by eqs.(1-2), using a 3D BSE-cum-SDE approach which is attuned to hadron
spectroscopy of both varieties simultaneously \cite{18,19}. The on-shell
value agrees with experiment \cite{27}, while the off-shell parameter
$\lambda$ is rather close to unity, signifying a change of sign
for $\theta(q^2)$ in just the transition region between  space-like
and time-like momenta.

One of us (ANM) wishes to thank Prof.W.Y.Pauchy Hwang and the Dept.of
Physics of National Taiwan University for kind hospitality, as well as
the National Science Council of R.O.C. for financial support.

\end{document}